\documentclass[aps,prd,twocolumn,superscriptaddress,showpacs]{revtex4}
\usepackage{graphicx}
\usepackage{epstopdf}
\usepackage{amsmath}
\usepackage{amsfonts}
\usepackage{amssymb}
\usepackage{latexsym}

\begin{document}
\title{The bosonized version of the Schwinger model in four dimensions: a blueprint for confinement?}
\author{Antonio Aurilia} \email{aaurilia@cpp.edu}
\affiliation{Department of Physics, California State Polytechnic University-Pomona, Pomona, California 91768, USA}
\author{Patricio Gaete} \email{patricio.gaete@usm.cl} 
\affiliation{Departmento de F\'{i}sica and Centro Cient\'{i}fico-Tecnol\'ogico de Valpara\'{i}so-CCTVal, 
Universidad T\'{e}cnica Federico Santa Mar\'{i}a, Valpara\'{i}so, Chile}
\author{Jos\'{e} A. Helay\"{e}l-Neto}\email{helayel@cbpf.br}
\affiliation{Centro Brasileiro de Pesquisas F\'{i}sicas (CBPF), Rio de Janeiro, RJ, Brasil}
\author{Euro Spallucci}\email{euro@ts.infn.it}
\affiliation{Dipartimento di Fisica Teorica, Universit\`a di Trieste and INFN, Sezione di Trieste, Italy}
\date{\today}

\begin{abstract}
For a $(3+1)$-dimensional generalization of the Schwinger model, we compute the interaction energy between two test charges. The result shows that the static potential profile contains a linear term leading to the confinement of probe charges, exactly as in the original model in two dimensions. We further show that the same 4-dimensional model  also appears as one version of the $ B \wedge F$ models in $(3+1)$ dimensions under dualization of Stueckelberg-like massive gauge theories. Interestingly, this particular model is characterized by the mixing between a $U(1)$ potential and an Abelian $3$-form field of the type that appears in the topological sector of QCD.
\end{abstract}
\maketitle

\section{Introduction}

It is generally agreed that two-dimensional field-theory models may provide an excellent and rich framework to test ideas in gauge theories. In fact, the interest in studying these models is basically connected to the possibility of obtaining exact solutions, which are believed to be shared by their more realistic counterparts in four dimensions.
Of these, the Schwinger model, also known as Quantum Electrodynamics in $(1+1)$-space-time dimensions, or ${QED}_2$\cite{Schwinger:1962tn,Coleman:1975pw} has probably enjoyed the greatest popularity due to some special features that it possesses. For example, the energy spectrum contains a massive mode in spite of the gauge invariance of the original Lagrangian, the charge is screened and confinement is enforced by the explicit occurrence of a rising Coulomb potential. To our mind, these special features represent the essential ingredients of a mechanism by which one hopes to understand the phenomenon of quark-binding into physical hadrons. These issues were first analyzed in ${QED}_2$ in Refs. \cite{Schwinger:1962tn,Schwinger:1962tp, Casher:1973uf,Coleman:1975pw}.

Unfortunately, against this suggestive two-dimensional perspective, it seems to us that a convincing analytical proof of color confinement in quantum chromodynamics (QCD) still eludes us. The root of the problem is well known: while asymptotic freedom is a well established property of the perturbative dynamics of QCD, the transition to infrared slavery is problematic because of non-perturbative effects that dominate in the large distance limit of the theory. Once this ?large distance limit? is defined in terms of some phenomenological scale of distance, the immediate problem is that of identifying the dynamical variables that operate in that limit. A hint about the nature of those hidden dynamical variables comes from the phenomenological bag models of hadrons: the partial success of those models indicate that, in the large distance limit of QCD, the spatial extension of hadrons and the bag degrees of freedom must somehow be included among those new dynamical variables. It is clear that, in order to speak meaningfully of a ?QCD-solution? of the confinement problem, one would expect that such variables should arise from the very dynamics of QCD and control the mechanisms of color confinement \cite{Luscher:1978rn}. 
This is where the extrapolation of results from two to four spacetime dimensions may play a significant role in the understanding of the confinement mechanism in QCD. For instance, the correspondence between the colorless topological sector of QCD and the zero-charge sector of ${QED}_2$ was noted long ago in Ref.\cite{Aurilia:1979dw} but never fully exploited;  The extrapolation from two to four dimensions, at least for the bosonized version of the Schwinger model, was considered in \cite{  Aurilia:1980jz} while a general "gauge mixing mechanism for the generation of mass" was proposed in \cite{Aurilia:1981xg}. 

Motivated by these observations, the general purpose of the present discussion is to communicate a deeper understanding  of the physical content of the $(3+1)$-dimensional generalization of the Schwinger model. The many avenues of research that are open to us were outlined in a research proposal by the authors \cite{Aurilia:2015qia}. However, it seems clear that the first line of inquiry is to explore in more detail the role of the Abelian $3$-form field among the physical observables of the model. It has long been known that this $3$-form field does not support any propagating degree of freedom, its sole physical effect consisting of a static interaction between two probe charges.This remarkable property is entirely analogous to the two dimensional case where in ${QED}_2$ there are no "photons" associated with the electromagnetic field \cite{Aurilia:1979dw}. Then, if the Schwinger model has any relevance in the issue of confinement in four dimensions, then the static potential induced by the Abelian $3$-form field must also exhibit the same behavior found in the two-dimensional case. We find that this reasonable expectation is fully supported by the explicit calculation of the interaction energy between two external test charges.

A second objective of this work is to elucidate the remarkable interplay between guge invariance and the appearance of mass in the physical spectrum of the Schwinger model. With hindsight, the emergence of this massive mode can be traced back directly to the dimensionality of the coupling constant in ${QED}_2$ which sets a mass scale in the model. Evidently this is not the case in ${QED}_4$ but a similar phenomenon takes place, at least in the bosonized version of the Schwinger model in $(3+1)$  dimensions. We illustrate how this same generalization of the S-model basically amounts to a Stueckelberg-like formulation of a massive gauge theory characterized by the mixing between a $U(1)$ potential and an Abelian $3$-form field.

Our work is organized according to the following outline: in Section II, we recall the salient features of dualization in terms of two simple Lagrangian systems and show their equivalence to different representations of a massive Proca field. In Section III,  using a path-integral approach, we compute the interaction energy, and hence the analytic form of the static potential in the bosonized version of the Scwinger model in four spacetime dimensions. Finally, some Concluding Remarks are cast in Sec. IV. 

Throughout the following discussion, the signature of the metric is ($+1,-1,-1,-1$).

\section{Dualization, gauge invariance and mass generation}

Let us start our considerations by recalling that the study of duality symmetry in gauge theories has been of considerable importance in 
order to provide an equivalent description of physical phenomena by distinct theories. As well-known, duality refers to a physical equivalence 
between two field theories which formulated in terms of different dynamical variables \cite{Hjelmeland:1997eg}. 

In order to put our discussion into context, we also recall that the dualization of Stueckelberg-like massive gauge theories 
and $ B \wedge F$ models follows from a general $p$ dualization of interacting theories in $d$ spacetime 
dimensions \cite{Ansoldi:1999wi,Smailagic:1999qw,Ansoldi:2000qs,Smailagic:2000hr,Smailagic:2001ch}. 
In particular, in the case of $(3+1)$ dimensions, the following $ B \wedge F$ models are found:
\begin{equation}
{\cal L}^{(1)} =  - \frac{1}{4}F_{\mu \nu }^2\left( A \right) + \frac{1}{{12}}H_{\mu \nu \rho }^2\left( B \right) 
+ \frac{m}{{24}}{\varepsilon ^{\mu \nu \rho \sigma }}{B_{\mu \nu }}{\partial _{[\rho }}{A_{\sigma ]}},  \label{Dual05}
\end{equation}

\begin{equation}
{\cal L}^{(2)} =  - \frac{1}{4}H_{\mu \nu }^2\left( B \right) + \frac{1}{{12}}F_{\mu \nu \rho }^2\left( A \right) 
+ \frac{m}{{24}}{\varepsilon ^{\mu \nu \rho \sigma }}{B_\mu }{\partial _{[\nu }}{A_{\rho \sigma ]}}, \label{Dual10}
\end{equation}

\begin{equation}
{\cal L}^{(3)} = \frac{1}{2}{\left( {{\partial _\mu }\varphi } \right)^2} - \frac{1}{{48}}F_{\mu \nu \rho \sigma }^2\left( A \right) 
+ \frac{m}{{24}}{\varepsilon ^{\mu \nu \rho \sigma }}\varphi {\partial _{[\mu }}{A_{\nu \rho \sigma ]}}. \label{Dual15}
\end{equation}

At this point, it is instructive to make a brief re-examination of equations (\ref{Dual05}) and (\ref{Dual10}). For this purpose, we observe 
that the Lagrangian density (\ref{Dual05}) may be rewritten as
\begin{equation}
{{\cal L}^{\left( 1 \right)}} =  - \frac{1}{4}F_{\mu \nu }^2 - \frac{1}{2}{\tilde H_\sigma }{\tilde H^\sigma } 
- \frac{m}{6}{\tilde H^\sigma }{A_\sigma }, \label{Dual25}
\end{equation}
where we have made use of ${\tilde H^\mu } = 
{\raise0.5ex\hbox{$\scriptstyle 1$}\kern-0.1em/\kern-0.15em\lower0.25ex\hbox{$\scriptstyle 2$}}{\varepsilon ^{\mu \nu \lambda \rho }}{\partial _\nu }{B_{\lambda \rho }}$.

Next, in order to eliminate the dual-field $H^{\sigma}$ care must be taken, for it satisfies the constraint 
${\partial _\mu }{\tilde H^\mu } = 0$ (Bianchi identity). Thus, to take into account the constraint, we shall introduce a Lagrange 
multiplier $\chi$. In such a case, the corresponding effective Lagrangian density (\ref{Dual25}) reads 
\begin{equation}
{{\cal L}^{\left( 1 \right)}} = - \frac{1}{4}F_{\mu \nu }^2 - \frac{1}{2}{\tilde H_\sigma }{\tilde H^\sigma }
- \frac{m}{6}{\tilde H^\sigma }{A_\sigma } + \chi {\partial _\sigma }{\tilde H^\sigma }   .\label{Dual30}
\end{equation}
By defining ${Z_\sigma } \equiv {A_\sigma } + \frac{6}{m}{\partial _\sigma }\chi$, with $  {Z_{\mu \nu }} 
= {F_{\mu \nu }}$, we readily verify that
\begin{equation}
{{\cal L}^{\left( 1 \right)}} =  - \frac{1}{4}Z_{\mu \nu }^2 - \frac{1}{2}{\tilde H_\sigma }{\tilde H^\sigma } 
- \frac{m}{6}{\tilde H^\sigma }{Z_\sigma }.\label{Dual35} 
\end{equation}
By a further definition of the fields, ${W_\sigma } \equiv {\tilde H_\sigma } + \frac{m}{6}{Z_\sigma }$, we find that the Lagrangian 
density (\ref{Dual05}) can be brought to the form
\begin{equation}
{{\cal L}^{\left( 1 \right)}} =  - \frac{1}{4}Z_{\mu \nu }^2 + \frac{1}{2}{\mu ^2}Z_\mu ^2, \label{Dual40}
\end{equation}
with ${\mu ^2} \equiv {\raise0.5ex\hbox{$\scriptstyle {{m^2}}$}
	\kern-0.1em/\kern-0.15em
	\lower0.25ex\hbox{$\scriptstyle {36}$}}$. We immediately see that the Lagrangian density (\ref{Dual40}) exhibits a Proca-type mass term.

We now turn our attention to the Lagrangian density (\ref{Dual10}). It is convenient to rewrite this equation in the alternative form 
\begin{equation}
{{\cal L}^{\left( 2 \right)}} = \frac{1}{4}\tilde H_{\mu \nu }^2 + \frac{1}{{12}}F_{\mu \nu \rho }^2 
+ \frac{m}{{24}}{\tilde H^{\rho \sigma }}{A_{\rho \sigma }}, \label{Dual45}
\end{equation}
where ${\tilde H^{\mu \nu }} = {\raise0.5ex\hbox{$\scriptstyle 1$}
	\kern-0.1em/\kern-0.15em
	\lower0.25ex\hbox{$\scriptstyle 2$}}{\varepsilon ^{\mu \nu \lambda \rho }}{H_{\lambda \rho }}$.

It is worthy to notice that the $B^{\mu}$- field appears only through ${\tilde H^{\mu\nu}}$.
Again, in order to eliminate the dual-field $\tilde H^{\mu\nu}$ care must be taken, for it satisfies the 
constraint ${\partial _\mu }{\tilde H^{\mu\nu} } = 0$. As before, we shall introduce a Lagrange multiplier $\chi_{\nu}$. 
It gives rise to the following Lagrangian density,
\begin{equation}
{{\cal L}^{\left( 2 \right)}} = \frac{1}{4}\tilde H_{\mu \nu }^2 + \frac{1}{{12}}F_{\mu \nu \rho }^2 
+ \frac{m}{{24}}{\tilde H^{\mu \nu }}{A_{\mu \nu }} - \frac{1}{2}{\tilde H^{\mu \nu }}{\chi _{\mu \nu }},\label{Dual50}
\end{equation}
where ${\chi _{\mu \nu }} = {\partial _\mu }{\chi _\nu } - {\partial _\nu }{\chi _\mu }$.
Now, letting ${Z_{\mu \nu }} = {A_{\mu \nu }} - \frac{{12}}{m}{\chi _{\mu \nu }}$, we obtain
\begin{equation}
{{\cal L}^{\left( 2 \right)}} = \frac{1}{4}\tilde H_{\mu \nu }^2 + \frac{1}{{12}}F_{\mu \nu \rho }^2 
+ \frac{m}{{24}}{{\tilde H}^{\mu \nu }}{Z_{\mu \nu }}. \label{Dual55}
\end{equation}
It should be further noted that, by defining ${W_{\mu \nu }} = {{\tilde H}_{\mu \nu }} 
+ \frac{m}{{12}}{Z_{\mu \nu }}$, equation (\ref{Dual55}) reduces to 
\begin{equation}
{{\cal L}^{\left( 2 \right)}} = \frac{1}{{12}}F_{\mu \nu \rho }^2 - \frac{1}{2}{\mu ^2}Z_{\mu \nu }^2, \label{Dual60} 
\end{equation}
where we have written, ${\mu ^2} = \frac{{{m^2}}}{{288}}$, and 
$F_{\mu \nu \rho }^2 = Z_{\mu \nu \rho }^2$. Thus ${{\cal L}^{\left( 2 \right)}}$ describes a massive field of spin $1$, exactly a 
Proca equation, although ${Z_{\mu \nu }} \in \left[ {\left( {1,0} \right) \oplus \left( {0,1} \right)} \right]$. Actually, a massive
rank-two skew-symmetric tensor field is, on-shell, equivalent to a Proca field.

In short, equations (\ref{Dual05}) and (\ref{Dual10}) are equivalent; both of these equations describe a Proca field.

Considering, finally, equation (\ref{Dual15}), we find that this model reduces to a  massless Schwinger model in $(3+1)$ dimensions, 
as we shall indicate it below.

\section{Interaction energy}

Inspired by the preceding observation, we shall now consider the $(3+1)$-dimensional generalization of the Schwinger model, as originally 
introduced in Ref.\cite{Aurilia:1979dw}. As we have already noticed, we will work out the static potential for this $(3+1)$ generalization, 
via a path-integral approach. To this end, we consider the bosonized form of the Schwinger model in D=$(3+1)$, that is,
\begin{equation}
{\cal L} = \frac{1}{2}{\left( {{\partial _\mu }\phi } \right)^2} + \frac{1}{2}m_\phi ^2{\phi ^2} 
+ \frac{g}{{6\sqrt \pi  }}{\partial _\mu }\phi\ {\varepsilon ^{\mu \nu \rho \sigma }}{A_{\nu \rho \sigma }} 
- \frac{1}{{48}}F_{\mu \nu \rho \sigma }^2, \label{Scwhin3-05}
\end{equation}
where $g$ is a coupling constant and $m_\phi$ refers to the mass of the scalar field $\phi$.

We readily verify that when, ${m_\phi } \to 0$,
equation (\ref{Scwhin3-05}) reduces to equation (\ref{Dual15}).

According to usual procedure, integrating out the $\phi$ field induces an effective theory for the $A_{\nu \rho \sigma }$ field.  
It is now important to recall that the $ A_{\nu \rho \sigma }$ field  can also be written as ${A_{\nu \rho \sigma }} 
= {\varepsilon _{\nu \rho \sigma \lambda }}{\partial ^\lambda }\xi$  \cite{Aurilia:2004cb, Aurilia:2004fz}, 
where $\xi$ refers to an another scalar field. This then leads to the following effective theory for the model under consideration:
\begin{equation}
{\cal L} = \frac{1}{2}\left[ {\xi \ \Delta \left( {1 + \frac{{{\raise0.7ex\hbox{${{g^2}}$} \!\mathord{\left/
						{\vphantom {{{g^2}} \pi }}\right.\kern-\nulldelimiterspace}
					\!\lower0.7ex\hbox{$\pi $}}}}{{\left( {\Delta  - m_\phi ^2} \right)}}} \right)\Delta \ \xi } \right], \label{Scwhin3-10}
\end{equation}
where $\Delta  = {\partial _\mu }{\partial ^\mu }$.

We are now ready to compute the interaction energy between static pointlike sources. We start off our analysis by writing down the functional 
generator of the Green's functions, that is,
\begin{equation}
Z\left[ J \right] = \exp \left( { - \frac{i}{2}\int {{d^4}x{d^4}yJ(x)D(x,y)J(y)} } \right), \label{Scwhin3-15}
\end{equation}
where, $D(x,y) = \int {\frac{{{d^4}k}}{{{{\left( {2\pi } \right)}^4}}}D(k){e^{ - ikx}}}$, is the propagator. In this case, the corresponding 
propagator is given by
\begin{equation}
D(k) = \left( {1 - \frac{{m_\phi ^2}}{{{{\cal M}^2}}}} \right)\frac{1}{{{k^2}\left( {{k^2} + {{\cal M}^2}} \right)}} 
+ \frac{{m_\phi ^2}}{{{{\cal M}^2}}}\frac{1}{{{k^4}}}, \label{Scwhin3-20}
\end{equation}
where ${{\cal M}^2} = m_\phi ^2 - {\raise0.5ex\hbox{$\scriptstyle {{g^2}}$}\kern-0.1em/\kern-0.15em
	\lower0.25ex\hbox{$\scriptstyle \pi $}}$.

By means of expression $Z = {e^{iW\left[ J \right]}}$ and employing Eq. (\ref{Scwhin3-15}), ${W\left[ J \right]}$ takes the form 
\begin{eqnarray}
W\left[ J \right] &=&  - \frac{1}{2}\int {\frac{{{d^4}k}}{{{{\left( {2\pi } \right)}^4}}}} {J^ * }\left( k \right)
\frac{{\left( {1 - \frac{{m_\phi ^2}}{{{{\cal M}^2}}}} \right)}}{{{k^2}\left( {{k^2} + {{\cal M}^2}} \right)}}
J\left( k \right) \nonumber\\
&-& \frac{1}{2}\int {\frac{{{d^4}k}}{{{{\left( {2\pi } \right)}^4}}}} {J^ * }\left( k \right)\frac{{m_\phi ^2}}{{{{\cal M}^2}}}
\frac{1}{{{k^4}}}J\left( k \right).
\label{Scwhin3-25}
\end{eqnarray}

Next, for $J({\bf x}) = \left[ {Q{\delta ^{\left( 3 \right)}}\left( {{\bf x} - {{\bf x}^{\left( 1 \right)}}} \right) 
	+ {Q^ \prime }{\delta ^{\left( 3 \right)}}\left( {{\bf x} - {{\bf x}^{\left( 2 \right)}}} \right)} \right]$, we obtain that the interaction 
energy of the system is given by
\begin{eqnarray}
V &=&  - Q{Q^ * }\int {\frac{{{d^3}k}}{{{{\left( {2\pi } \right)}^3}}}} 
\frac{{\left( {\frac{{{\raise0.5ex\hbox{$\scriptstyle {{g^2}}$}
						\kern-0.1em/\kern-0.15em
						\lower0.25ex\hbox{$\scriptstyle \pi $}}}}{{{\raise0.5ex\hbox{$\scriptstyle {{g^2}}$}
						\kern-0.1em/\kern-0.15em
						\lower0.25ex\hbox{$\scriptstyle \pi $}} - m_\phi ^2}}} \right)}}{{\left( {{{\bf k}^2} + {\raise0.5ex\hbox{$\scriptstyle {{g^2}}$}
				\kern-0.1em/\kern-0.15em
				\lower0.25ex\hbox{$\scriptstyle \pi $}} - m_\phi ^2} \right)}}
{e^{i{\bf k} \cdot {\bf r}}} \nonumber\\
&+& Q{Q^ * }\int {\frac{{{d^3}k}}{{{{\left( {2\pi } \right)}^3}}}} \left( {\frac{{m_\phi ^2}}{{{\raise0.5ex\hbox{$\scriptstyle {{g^2}}$}
				\kern-0.1em/\kern-0.15em
				\lower0.25ex\hbox{$\scriptstyle \pi $}} - m_\phi ^2}}} \right)\frac{1}{{{{\bf k}^4}}}{e^{i{\bf k} \cdot {\bf r}}}, \label{Scwhin3-30}
\end{eqnarray}
where ${\bf r} = {{\bf x}^{\left( 1 \right)}} - {{\bf x}^{\left( 2 \right)}}$.

This, together with ${Q^ \prime }=-Q$, yields finally
\begin{eqnarray}
V &=& \frac{{{Q^2}}}{{4\pi }}\frac{{{\raise0.5ex\hbox{$\scriptstyle {{g^2}}$}
			\kern-0.1em/\kern-0.15em
			\lower0.25ex\hbox{$\scriptstyle \pi $}}}}{{{{\left( {{\raise0.5ex\hbox{$\scriptstyle {{g^2}}$}
						\kern-0.1em/\kern-0.15em
						\lower0.25ex\hbox{$\scriptstyle \pi $}} - m_\phi ^2} \right)}^2}}}\frac{1}{L}\left( {1 - {e^{ - \sqrt {{\raise0.5ex\hbox{$\scriptstyle {{g^2}}$}
					\kern-0.1em/\kern-0.15em
					\lower0.25ex\hbox{$\scriptstyle \pi $}} - m_\phi ^2} L}}} \right)  \nonumber\\
&+& \frac{{{Q^2}}}{{4\pi }}\frac{{m_\phi ^2}}{{2\left( {{\raise0.5ex\hbox{$\scriptstyle {{g^2}}$}
				\kern-0.1em/\kern-0.15em
				\lower0.25ex\hbox{$\scriptstyle \pi $}} - m_\phi ^2} \right)}}L, \label{Scwhin3-35}
\end{eqnarray}
where $L = |{\bf r}|$. One immediately sees that the above static potential profile is analogous to that encountered in the two-dimensional 
Schwinger model. Incidentally, in order to put our discussion into context it is useful to summarize the relevant aspects of the 
two-dimensional Schwinger model. In such a case, we begin by recalling the bosonized form of the model under consideration \cite{Gross:1995bp}:
\begin{eqnarray}
{\cal L} &=&  - \frac{1}{4}F_{\mu \nu }^2 + \frac{1}{2}{\left( {{\partial _\mu }\phi } \right)^2} 
- \frac{e}{{2\sqrt \pi  }}{\varepsilon ^{\mu \nu }}{F_{\mu \nu }}\phi \nonumber\\
&+&m\sum \left( {\cos \left( {2\pi \phi  + \theta } \right) - 1} \right), \label{Scwhin3-40}
\end{eqnarray}
where $\sum  = \left( {\frac{e}{{2{\pi ^{{\raise0.5ex\hbox{$\scriptstyle 3$}
						\kern-0.1em/\kern-0.15em
						\lower0.25ex\hbox{$\scriptstyle 2$}}}}}}} \right){e^{{\gamma _E}}}$ with ${\gamma _E}$ the Euler-Mascheroni constant and $\theta$ refers to 
the $\theta$-vacuum. 

Consequently, by using the gauge-invariant but path-dependent variables formalism which provides a physically-based alternative to the Wilson 
loop approach \cite{Gaete:1999iy, Gaete:2001wh}, the static potential reduces to
\begin{equation}
V = \frac{{{Q^2}}}{2}\frac{{\sqrt \pi  }}{e}\left( {1 - {e^{ - \frac{e}{{\sqrt \pi  }}L}}} \right), \label{Scwhin3-45}
\end{equation}
for the massless case. On the other hand, for the massive case ($\theta=0$), the static potential then becomes
\begin{equation}
V = \frac{{{Q^2}}}{{2\lambda }}\left( {1 + \frac{{4\pi m\sum }}{{{\lambda ^2}}}} \right)\left( {1 - {e^{ - \lambda L}}} \right) 
+ \frac{{{q^2}}}{2}\left( {1 - \frac{{{\raise0.5ex\hbox{$\scriptstyle {{e^2}}$}
				\kern-0.1em/\kern-0.15em
				\lower0.25ex\hbox{$\scriptstyle \pi $}}}}{{{\lambda ^2}}}} \right)L, \label{Scwhin3-50}
\end{equation}
where ${\lambda ^2} = \frac{{{e^2}}}{\pi } + 4\pi m\sum$. The above results clearly show that the $(3+1)$-D generalization of the Schwinger 
model is structurally identical to the $(1+1)$-D Schwinger model.

In this perspective it is worth recalling that there is an alternative way of obtaining the Lagrangian density  (\ref{Scwhin3-10}), 
which provides a complementary view into the physics of confinement. In fact, we refer to a theory of antisymmetric tensor fields that 
results from the condensation of topological defects as a consequence of the Julia-Toulouse mechanism. 
More precisely, the Julia-Toulouse mechanism is a condensation process dual to the Higgs mechanism proposed in \cite{Quevedo:1996uu}. 
This mechanism describes phenomenologically the electromagnetic behavior of antisymmetric tensors in the presence of magnetic-branes 
(topological defects) that eventually condensate due to thermal and quantum fluctuations. Using this phenomenology we have discussed 
in \cite {Gaete:2004dn,Gaete:2005am} the dynamics of the extended charges (p-branes) inside the new vacuum provided by the condensate. 
Actually, in \cite {Gaete:2004dn} we have considered the topological defects coupled both longitudinally and 
transversally to two different tensor potentials, $A_p$ and $B_q$, such that $p+q+2=D$, where $D=d+1$ space-time dimensions.

We skip all the technical details and refer to \cite{Gaete:2004dn} for them. Thus, after the condensation, the Lagrangian density turns out to be
\begin{eqnarray}
{\cal L} &=& \frac{{{{\left( { - 1} \right)}^q}}}{{2\left( {q + 1} \right)!}}{\left[ {{H_{q + 1}}\left( {{B_q}} \right)} \right]^2} 
+ e{B_q}{\varepsilon ^{q,\alpha ,p + 1}}{\partial _\alpha }{\Lambda _{p + 1}} \nonumber\\
&+& \frac{{{{\left( { - 1} \right)}^{p + 1}}}}{{2\left( {p + 2} \right)!}}{\left[ {{F_{p + 2}}\left( {{\Lambda _{p + 1}}} \right)} \right]^2} 
\nonumber\\
&+& \frac{{{{\left( { - 1} \right)}^{p + 1}}\left( {p + 1} \right)!}}{2}{m^2}\Lambda _{p + 1}^2,  \label{Scwhin3-55}
\end{eqnarray}
showing a $B$$\wedge$$F$ type of coupling between the $B_q$ potential with the tensor $\Lambda_{p+1}$ carrying the degrees of freedom of the 
condensate. Following our earlier procedure \cite{Gaete:2004dn}, the effective theory that results from integrating out the fields representing 
the vacuum condensate, is given by
\begin{equation}
{\cal L} = \frac{{{{\left( { - 1} \right)}^{q + 1}}}}{{2\left( {q + 1} \right)!}}{H_{q + 1}}\left( {{B_q}} \right)
\left( {1 + \frac{{{e^2}}}{{\Delta  - {m^2}}}} \right){H^{q + 1}}\left( {{B_q}} \right). \nonumber\\
\label{Scwhin3-60}
\end{equation}  
Hence we see that this expression with $p=-1$ and $q=3$ becomes
\begin{equation}
{\cal L} = \frac{1}{{2 \times 4!}}{F_{\mu \nu \rho \lambda }}\left( A \right)\left( {1 + \frac{{{e^2}}}{{\Delta  - {m^2}}}} \right)
{F^{\mu \nu \rho \lambda }}\left( A \right). \label{Scwhin3-65}
\end{equation}
It is straightforward to verify that Eq. (\ref{Scwhin3-65}) reduces to Eq. (\ref{Scwhin3-10}).

In this way, we establish a new connection among different effective theories. It must be clear from this discussion that the above connections 
are of interest from the point of view of providing unifications among diverse models as well as exploiting their equivalence in explicit 
calculations.

\section{Concluding Remarks}

Finally, the point we wish to emphasize is that there are two generic features that are common in the four-dimensional case and their 
upper/lower extensions, as we shall show below. First, the existence of a linear potential, leading to the confinement of static charges. 
The second point is related to the correspondence among diverse effective theories. To see this, it should be noted that by using the 
methodology illustrated in \cite{  Smailagic:1999qw}, we have that one of the $ B \wedge F$ models in $(4+1)$ dimensions is given by the 
mixing between a $U(1)$ potential and an Abelian $3$-form field by means of a topological mass term, that is,
\begin{eqnarray}
{\cal L}^{\left( {4 + 1} \right)} &=&  - \frac{1}{4}{F_{\mu \nu }}\left( A \right){F^{\mu \nu }}\left( A \right) 
+ \alpha {H_{\mu \nu \kappa \lambda }}\left( C \right){H^{\mu \nu \kappa \lambda }}\left( C \right) \nonumber\\
&+& \beta {\varepsilon ^{\mu \nu \kappa \lambda \rho }}{A_\mu }{\partial _\nu }{C_{\kappa \lambda \rho }}, \label{CR-05}
\end{eqnarray} 
with $\alpha  =  - \frac{1}{{48}}$ and $ \beta  = \frac{\sigma }{6}$, where the parameter $\beta$ has mass dimension. This model was considered 
in \cite{Cocuroci:2013bga}, and the main motivation to consider this model is based on the possible connection with dark energy.

However, we shall start from the five-dimensional spacetime model
\begin{eqnarray}
{\cal L}^{\left( {4 + 1} \right)} &=&  - \frac{1}{4}{F_{\hat \mu \hat \nu }}{F^{\hat \mu \hat \nu }} 
+ \alpha {H_{\hat \mu \hat \nu \hat \kappa \hat \lambda }}{H^{\hat \mu \hat \nu \hat \kappa \hat \lambda }} \nonumber\\
&+& \beta {\varepsilon ^{\hat \mu \hat \nu \hat \kappa \hat \lambda \hat \rho }}{A_\mu }{\partial _\nu }{C_{\hat \kappa \hat \lambda \hat \rho }} + \frac{1}{{12}}m_C^2{C_{\hat \mu \hat \nu \hat \rho }}{C^{\hat \mu \hat \nu \hat \rho }}, \nonumber\\
\label{CR-10}
\end{eqnarray}
with the additional presence of a mass term $m_C$ for the Abelian $3$-form field; this explicit mass term makes a difference: if it were not introduced, the model
could be reduced to nothing but a Proca-type model in $(4+1)$ dimensions. Next, we perform its dimensional reduction along the 
lines of \cite{Cocuroci:2013bga,Gaete:2012yu}: 
${A^{\hat \mu }} \to \left( {{A^{\bar \mu }},{A^4}} \right)$, ${A^4} = \phi$, ${\partial _4}\left( {everything} \right) = 0$, 
${C^{\hat \mu \hat \nu \hat \kappa }} = \left( {{C^{\bar \mu \bar \nu \bar \kappa }},{C^{\bar \mu \bar \nu 4}}} \right)$ 
and ${C^{\bar \mu \bar \nu 4}} = {B^{\bar \mu \bar \nu }}$. Carrying out this prescription in equation (\ref{CR-10}), we then obtain
\begin{eqnarray} 
{{\cal L}^{\left( {3 + 1} \right)}} &=&  - \frac{1}{4}{F_{\bar \mu \bar \nu }}{F^{\mu \nu }} 
+ \frac{1}{2}{\left( {{\partial _{\bar \mu} }\phi } \right)^2} + \alpha {H_{\bar \mu \bar \nu \bar \kappa \bar \lambda }}
{H^{\bar \mu \bar \nu \bar \kappa \bar \lambda }} \nonumber\\
&-& 4\alpha {G_{\bar \mu \bar \nu \bar \kappa }}{G^{\bar \mu \bar \nu \bar \kappa }} - 3\beta {\varepsilon ^{4 \bar \mu \bar \nu \bar \kappa 
		\bar \lambda }}{A_{\bar \mu} }{\partial _{\bar \nu} }{B_{\bar \kappa \bar \lambda }} \nonumber\\
&-& \beta {\varepsilon ^{4 \bar \nu \bar \kappa \bar \lambda \bar \rho }}\phi {\partial _{\bar \nu} }{C_{\bar \kappa \bar \lambda \bar \rho }} 
+ \frac{{m_C^2}}{{12}}{C_{\bar \mu \bar \nu \bar \rho }}{C^{\bar \mu \bar \nu \bar \rho }} \nonumber\\
&-& \frac{{m_C^2}}{4}{B_{\bar \mu \bar \nu }}{B^{\bar \mu \bar \nu }}, \label{CR-15}
\end{eqnarray}
where $\bar \mu ,\bar \nu ,\bar \kappa ,\bar \lambda ,\bar \rho  = 0,1,2,3$. Making use of an additional dimensional reduction, that is, 
$ {A^{\bar \mu }} \to \left( {{A^\mu },{A^3}} \right)$, ${\partial _3}\left( {everything} \right) = 0$, ${B^{\bar \mu \bar \nu }} 
= \left( {{B^{\mu \nu }},{C^{\mu }}} \right)$
\begin{eqnarray}
{{\cal L}^{\left( {2 + 1} \right)}} &=&  - \frac{1}{4}{F_{\mu \nu }}{F^{\mu \nu }} + 12\alpha {G_{\mu \nu }}{G^{\mu \nu }} \nonumber\\
&-& 6\beta {\varepsilon ^{\mu \nu \kappa }}{A_\mu }{\partial _\nu }{C_\kappa } 
+ \frac{{m_C^2}}{2}{C_\mu }{C^\mu}, \label{CR-20}
\end{eqnarray}
where ${G_{\mu \nu }} = {\partial _\mu }{C_\nu } - {\partial _\nu }{C_\mu }$. Next, after performing the integration over $C_{\mu}$, 
the induced effective Lagrangian density is given by
\begin{equation}
{{\cal L}^{\left( {2 + 1} \right)}} =  - \frac{1}{4}{F_{\mu \nu }}\left( {1 + \frac{\sigma }{{\left( {\Delta  + m_C^2} \right)}}} \right)
{F^{\mu \nu }}. \label{CR-25}
\end{equation}
Again, by applying the gauge-invariant formalism, the corresponding static potential for two opposite charges located at ${\bf y}$ and 
${\bf y}\prime$ turns out to be
\begin{equation}
V =  - \frac{{{q^2}}}{{2\pi }}{K_0}\left( {ML} \right) + \frac{{{q^2}m_C^2}}{{4M}}L, \label {CR-30}
\end{equation}
where $L = |{\bf y} - {{\bf y}^ \prime }|$ and ${M^2} = {\sigma ^2} + m_C^2$. In summary, then: this potential displays the conventional 
screening part, encoded in the Bessel function, and the linear confining potential. As expected, confinement disappears whenever ${m_C} \to 0$
and also in the case $m_{C}$ is non-trivial, but much smaller than the topological mass parameter, $\sigma$.

A final consideration we would like to raise concerns the presence of some sort 
of fundamental mechanism that endows one of the gauge potentials, the $p$- or the
$(p+1)$-form, with a Proca-type mass term: if only the usual field-strength squared and
the topological mass terms are present, a field reshuffling is always possible to be done 
and one of the gauge potentials can be integrated over yielding, at the end, a Proca-like 
$p$-form or $(p+1)$-form massive model; exactly like we have worked out for the Lagrangians
(\ref{Dual05}) and (\ref{Dual10}). However, if a more fundamental mechanism is at work
(like the Higgs mechanism, for example) that gives an explicit (non-topological) mass term
to one of the gauge fields, then the simple equivalence to a $p$-form
Proca field is no longer true and a confining contribution to the static interparticle potential
shows. We would like to conclude our work by pointing out the relationship between the
generation of a non-topological mass and the confinig profile of the interparticle potential.

\section{Acknowledgments}

One of us (P. G.) wishes to thank the Abdus Salam ICTP for hospitality, the Field Theory Group of the COSMO/CBPF for the pleasant visit with the PCI-BEV/MCTIC support. P. G. was partially supported by Fondecyt (Chile) grant 1130426 and by Proyecto Basal FB0821.

\end{document}